\documentclass[11pt]{article}
\pdfoutput=1
\usepackage{jheppub}
\usepackage{pstool}
\pdfoutput=1
\usepackage{color}
\usepackage{float}
\usepackage{array}
\usepackage{amsthm}
\usepackage[labelsep=quad]{subcaption}
\usepackage{epstopdf}
\usepackage{placeins}
	
\usepackage{epsfig}

\def\pb[#1,#2]{\{#1, #2\}}
\def\deb[#1,#2]{[#1,#2]_{\text{D.B.}}}

\def\Or[#1]{{\text{O}}\left({#1}\right)}
\def\dotl[#1,#2]{\left\langle #1,\, #2 \right\rangle}
\def\dotlb[#1,#2]{\left\langle #1,\, #2 \right\rangle}
\def\dotlm[#1,#2]{\left[ #1,\, #2 \right]}
\def\dotp[#1,#2]{(\vect{#1} \cdot\vect{#2})}
\def\aff[#1,#2]{\hat{#1}(#2)}

\def\n4sym{{\cal N}=4 SYM}
\def\>{\rangle}
\def\<{\langle}
\def\weight[#1,#2,#3]{\{(#1),#2,#3\}}
\def\ads[#1]{$\text{AdS}_{#1}$}

\newcommand{\be}{\begin{equation}}
\newcommand{\ee}{\end{equation}}
\newcommand{\ba}{\begin{align}}
\newcommand{\ea}{\end{align}}
\newcommand{\bs}{\begin{split}}
	\def\sess\end{split}
\newcommand{\vect}[1]{{\boldsymbol{#1}}}

\def \bea {\begin{eqnarray}}
\def \eea {\end{eqnarray}}
\def \bea* {\begin{eqnarray*}}
	\def \eea* {\end{eqnarray*}}

\def \be {\begin{equation}}
\def \ee {\end{equation}}
\def \bes {\begin{equation*}}
\def \ees {\end{equation*}}

\title{The Large-$c$ Virasoro Identity Block is a Semi-Classical Liouville Correlator}
\author[a]{Gideon Vos,}
\emailAdd{vos@fzu.cz}
\affiliation[a]{Central European Institute for Cosmology, \\ FZU, Na Slovance 1999/2, 182 21 Prague 8, Czech Republic\\}

\date{}

\abstract{It will be shown analytically that the light sector of the identity block of a mixed heavy-light correlator in the large central charge limit is given by a correlation function of light operators on an effective background geometry. This geometry is generated by the presence of the heavy operators. It is shown that this background geometry is a solution to the Liouville equation of motion sourced by corresponding heavy vertex operators and subsequently that the light sector of the identity block matches the Liouville correlation function in the semi-classical limit. This method effectively captures the spirit of Einstein gravity as a theory of dynamical geometry in AdS/CFT. The reason being that Liouville theory is closely related to semi-classical asymptotically AdS$_3$ gravity.} 
\begin{document}
	\maketitle

\section{Introduction}	
Since the very beginning of its construction it was known that two-dimensional conformal field theory is closely intertwined with the geometry of punctured Riemann surfaces. Examples include both its start as a method for studying the Liouville model, up to it's realization in worldsheet string theory. The object of this paper will be to further strengthen this connection by demonstrating that in certain regimes operators with large scaling dimensions have the effect of generating an effective geometry upon which the dynamics of the light operators takes place.

In particular the object of study of this paper will be the Virasoro identity block in the semi-classical large central charge limit. Within the large central charge limit, based on their scaling weight, one typically distinguishes between heavy and light operators. The object under consideration will be correlators of the schematic form H...HLL, i.e. two light operators and a generic number of heavy operators. The goal will be to demonstrate analytically that the light operator dependence of the identity block can be interpreted as a correlation function of the light operators on top of an effective geometry generated by the heavy operators.

Furthermore, it will be demonstrated that this geometry is the same as the one generated by heavy vertex operators in Liouville theory. This remarkable equivalence ultimately stems from the fact that the accessory parameters of both relevant stress tensors are either fixed by the monodromy method for semi-classical CFT or by demanding by demanding that the Weyl factor in front of the metric is positive definite and single-valued in the case of Liouville theory. It was shown that when the monodromy method is used to compute the identity block these two methods coincide \cite{Hulik:2016ifr}.

In both cases the geometry in question is the metric that uniformizes the punctured Riemann sphere with weights at the marked points that correspond to the ratio of the heavy scaling weights and the central charge. This resulting geometry has constant negative curvature everywhere except at the locations of the operator insertions. The proof provided in this text is essentially a generalized analytic version of the algebraic proof of \cite{Fitzpatrick:2015zha} for HHLL correlators. In the process we will see that in this regime these generic correlators compute corresponding vertex operator correlators in semi-classical Liouville correlators, the holographic implications of this result are discussed at the end. 

The employed method involves allowing an extension to the set of conformal transformations from single-valued holomorphic transformations to multi-valued functions that are holomorphic everywhere except for a finite number of branch points. This will be justified by showing that the transformation preserves the general meromorphic structure of the stress tensor expectation value as expected from the conformal Ward identity. The result is also justified heuristically by providing a set of sensible physical results \cite{Fitzpatrick:2015zha,Banerjee:2018tut,Vos:2018vwv,Anous:2019yku}.

It is well-known that Liouville theory, at the classical level, is equivalent to Einstein gravity on an asymptotically AdS$_3$ background \cite{Coussaert:1995zp}. Furthermore the connection between Riemann surfaces, 2d CFT and/or 3d gravity has been carefully investigated in the past. For instance, either through the quantization of Teichm\"uller space \cite{Verlinde:1989ua}, or by studying the partition function on arbitrary genus Riemann surfaces \cite{Friedan:1986ua}. The structure constants were computed before by constructing the identity cross-channel from a universal crossing kernel and were found to approach Liouville theory structure constants in the large-$c$ limit \cite{Collier:2019weq}. Finally, the link between quantum Liouville theory and uniformization was conjectured by Polyakov and proven in \cite{TZ, Takhtajan:2001uj}. The main advantage of the method of this paper is that, beyond deriving the equations of asymptotically AdS$_3$ from semi-classical CFT, it also captures the spirit of Einstein gravity. The heavy operators inserted in a correlator generate a geometry upon which light probe operators move around, or as Wheeler encapsulated it: ``Space tells matter how to move, matter tells space how to curve.''

On another note, in the past it was argued that quantizing Liouville theory is perhaps not the way forward to obtain a quantum theory of gravity. That is, in the sense that quantizing Liouville is Liouville is akin to trying to obtain statistical mechanics of a system by quantizing it's coarse-grained thermodynamic degrees of freedom \cite{Martinec:1998wm}. In this paper I will sharpen this statement by arguing that the appearance of a Liouville regime is is a generic feature of gapped, unitary, semi-classical CFTs. This is due to the fact that, as long as the theory is gapped, kinematical regimes exist where the identity block dominates correlation functions.

Finally I will comment on some holographic interpretations in particular the AdS$_3$ interpretation for the emergence of a universal Liouville sector and revisit the well-studied HHLL example with an aim at focusing on the geometric viewpoint.

\section{Probes on heavy states (space tells matter how to move...)}
In the large central charge limit we can distinguish between heavy and light operators based on how their scaling weights compare to $c$ in the large $c$ limit:
\begin{align}
& \text{Heavy:} \;\; O(x_i) \;\;\; \lim_{c\rightarrow \infty} \frac{H}{c} \propto \mathcal{O}(1),\\
& \text{Light:} \;\;\; Q(z_i) \;\;\; \lim_{c\rightarrow \infty} \frac{h}{c} = 0.
\end{align}
Object of this section is the study of mixed heavy-light identity block of correlators of the form H...HLL. The result will be familiar to readers of \cite{Fitzpatrick:2015zha} except the method presented here evades the algebraic need to construct conformal blocks out of products of transformed Virasoro generators. The method will be even more familiar to readers of \cite{Anous:2019yku}, the main differences between this and those two works is that this method will consider H...HLL correlators rather than HHLL correlators and this method emphasizes that it is a second-order pole cancellation property of the uniformizing geometry that plays a key role rather than the reduction of the stress-tensor expectation value from an $\mathcal{O}(c)$ to an $\mathcal{O}(c^0)$ meromorphic function. The main difference though is that this method exposes the underlying geometrical meaning of the uniformizing geometry which will be made clear once we compare notes with semi-classical Liouville theory in the next section.

The argument in this section can be summarized along the following lines:
\begin{itemize}
\item It will be assumed that a particular factorization property holds for large-$c$ conformal blocks; the heavy-light decoupling property of semi-classical conformal blocks into a heavy and light sector \cite{recursion}.
\item A  (multi-valued) conformal transformation, one that uniformizes the punctured Riemann sheet, will be applied to this correlator. This has the effect of subtracting out the second-order poles at the heavy operator insertions and part of the first-order poles.
\item It will be shown that this transformation preserves the meromorphic structure of the stress-tensor. It will be assumed that in this new conformal frame the Virasoro Ward identity still holds. In particular, all residues of the first order poles will be collected and identified with the derivatives of the correlation function with respect to spatial points in the uniformizing frame coordinates.
\item This will show that this conformal frame has the particular property that under subsequent conformal transformations the heavy operator dependence transforms trivially whereas the light operators pick up the usual CFT Jacobian factors.
\item We will exploit the fact that the uniformizing property of the new conformal frame is preserved under global conformal transformations. This residual conformal symmetry will be used to fix the light operator dependence of the correlator in the new frame.
\end{itemize}

\subsection{Heavy-light factorization and multi-valued conformal transformations}
It is known that in the large-$c$ limit conformal blocks of heavy operators take on an exponential form \cite{Fitzpatrick:2014vua,recursion}
\begin{equation}
\langle O_1(x_1)...O_n(x_n)\rangle \sim e^{-\frac{c}{6}f(x_i,h_i)},
\end{equation}
where $f(x_i,h_i)$ takes on $\mathcal{O}(c^0)$ values across its domain. The light operator decoupling property tells us that the following related mixed correlator is given by
\begin{equation}
\langle Q(z_1)Q(z_2)O_1(x_1)...O_n(x_n)\rangle \sim Q(z_i,x_i)e^{-\frac{c}{6}f},
\end{equation}
here $Q(z_i,x_i)$ is a $\mathcal{O}(c^0)$ function as well, but now note that it is a function of both the light \textit{and} heavy operators. For both of these cases the effect of inserting a stress tensor in the correlator is fixed by the Virasoro Ward identity 
\begin{equation}
\langle T(z) O...O\rangle = \left( \sum_i \frac{H_i}{(z-x_i)^2} - \frac{\frac{c}{6}\partial_i f}{z-x_i}\right) e^{-\frac{c}{6}f} \equiv T(z) e^{-\frac{c}{6}f},
\end{equation}
note that this expectation value is $\mathcal{O}(c)$. Similarly
\begin{align}
&\langle T(z) QQO...O\rangle = \left( \sum_Q \frac{h_i}{(z-z_i)^2} + \frac{1}{z-z_i}\frac{\partial_{z_i} Q}{Q}\right)Qe^{-\frac{c}{6}f} \nonumber \\
& + \left( \sum_H \frac{H_i}{(z-x_i)^2} -\frac{\frac{c}{6}\partial_if}{z-x_i}+\frac{1}{z-x_i}\frac{\partial_{x_i} Q}{Q}\right) Qe^{-\frac{c}{6}f} \equiv  T_Q(z)Qe^{-\frac{c}{6}f},
\end{align}
which is also of order $\mathcal{O}(c)$ but only due to the presence of the terms that also occur in $T(z)$. Define the conformal transformation $u(z)$ by the following property
\begin{equation}
\left( \frac{dz}{du}\right)^2 T(z(u)) + \frac{c}{12}S[z,u]=0,
\label{inverseschwarzianode}
\end{equation}
where $S[z,u]$ is the Schwarzian derivative given by
\begin{equation}
S[z,u] = \frac{z'''(u)}{z'(u)} - \frac{3}{2}\left(\frac{z''(u)}{z'(u)}\right)^2.
\end{equation}
This conformal frame has the property that it exploits the inhomogeneous nature of the stress tensor transformation rule to subtract\footnote{This has the effect of rendering the stress tensor $\mathcal{O}(c^0)$, this was exploited to great effect in \cite{Fitzpatrick:2015zha,Anous:2019yku}, but will not play a role here other than consistency in powers of $c$.} out the first two terms from the heavy sector contribution to $T_Q(z)$. This procedure is not entirely well-defined though as we will see the function $u(z)$ will contain branch points, which means that $u(z)$ does not admit a Laurent expansion everywhere, which technically disqualifies it as a conformal transformation. Pushing ahead and performing this transformation anyway leads to
\begin{equation}
T_Q(u) = \left(\frac{dz}{du}\right)^2 \left( \sum_{i}^{2} \frac{h_i}{(z(u)-z_i)^2} + \frac{G_{z_i}}{z(u)-z_i}+ \sum_{i}^{n}\frac{G_{x_i}}{z(u)-x_i}\right).
\label{uniformizedT}
\end{equation}
Here the new accessory parameters $G$ are defined through
\begin{equation}
G_{z_i}=\frac{\partial}{\partial z_i} \log(Q), \hspace{5mm} G_{x_i}=\frac{\partial}{\partial x_i} \log(Q),
\end{equation}
This is a rather peculiar form for a stress tensor, it is uncommon to see a first-order pole without an accompanying second-order pole, the implications of this will be studied below.


\subsection{Implications arising from the light sector accessory parameters}
Away from the regular singular points at the heavy operator insertions we are essentially just in a different conformal frame. As such the Virasoro Ward identity should locally continue to hold. This suggests that the pole structure should be maintained and the residues at the first order poles at the location of the probe operators should give us the derivatives with respect to the probe insertion locations of the correlator in the new frame. This argument is heavily inspired by the analysis presented in \cite{Anous:2019yku} regarding heavy energy eigenstates.

To collect the residue of the first order pole we can exploit the fact that the Virasoro Ward identity imposes that $T_Q(u)$ has at most second order poles at $u_i \equiv u(z_i)$ and the first order poles of the probes are located at the same points as the double poles. Hence the following expression will generate the desired residue
\begin{equation}
R_i = \lim_{u\rightarrow u_i} \partial_u \, (u-u_i)^2 T_Q(u).
\end{equation}
After ignoring the terms that will not contribute, this approach gives us the following intermediate result
\begin{align}
& R_i = \lim_{u\rightarrow u_i} 2\left(\frac{\partial z}{\partial u}\right)\frac{\partial^2 z}{\partial u^2} \left( \frac{h_i (u-u_i)^2}{(z(u)-z_i)^2} + \frac{G_{z_i}(u-u_i)^2}{z(u)-z_i}\right)\\
& + \left(\frac{\partial z}{\partial u}\right)^2 \partial_u \left( \frac{h_i (u-u_i)^2}{(z(u)-z_i)^2} + \frac{G_{z_i}(u-u_i)^2}{z(u)-z_i}\right).
\end{align}
This expression can be considered as consisting of four terms, the second term is killed by the limit $u\rightarrow u_i$. After expanding $z(u)-z_i$ as
\begin{equation}
z(u)-z_i = \sum_{n=1} \frac{1}{n!}\frac{\partial^n z}{\partial u_i^n} (u-u_i)^n,
\label{seriesexp}
\end{equation}
Remember that the uniformizing coordinate only `knows' about the locations of the heavy operators, the probes are located at generic points hence we don't have to worry about non-generic behavior such as vanishing terms or non-analyticity in the expansion. The first and fourth term are simple and are respectively given by
\begin{equation}
2 h_I \lim_{u\rightarrow u_i} \left(\frac{\partial z}{\partial u}\right) \frac{\partial^2 z}{\partial u^2} \frac{(u-u_i)^2}{(\sum_{n=1} \frac{1}{n!}\frac{\partial^n z}{\partial u_i^n} (u-u_i)^n)^2} = 2 h_i \partial_{u_i} \log \frac{\partial z}{\partial u_i}
\end{equation} 
and
\begin{equation}
\lim_{u\rightarrow u_i} \partial_u \frac{G_{z_i} (u-u_i)^2}{z(u)-z_i}  = \left(\frac{\partial z}{\partial u_i}\right)^{-1} G_{z_i}.
\end{equation}
What makes the third term different is that it very naively appears to blow up, this is not true as it is simply the derivative of a quotient that is well-behaved in the appropriate limit. This is furthermore good as a greater-than-second order pole would be inconsistent with the Virasoro Ward identity. As we will see the singularities cancel among each other. This also makes this term tricky as it is the `next-to-leading order term' that survives the limit $u\rightarrow u_i$. First we expand the quotient
\begin{align}
& \lim_{u\rightarrow u_i} \left(\frac{\partial z}{\partial u}\right)^2 \partial_u \frac{h_i (u-u_i)^2}{(z(u)-z_i)^2} \nonumber \\
& =\lim_{u\rightarrow u_i} h_i \left(\frac{\partial z}{\partial u}\right)^2 {\partial u} \; (u-u_i)^2 \frac{1}{\left(\frac{\partial z}{\partial u}|_{u_i}\right)^2 (u-u_i)^2}\left(1-\frac{\frac{\partial^2z}{\partial u^2}|_{u_i}}{\frac{\partial z}{\partial u}_{u_i}}(u-u_i) + \mathcal{O}((u-u_i)^2)\right) \nonumber \\
& =-h_i \frac{\partial^2 z}{\partial u^2}|_{u_i} \left(\frac{\partial z}{\partial u}|_{u_i}\right)^{-3}.
\end{align}
This ends up bringing all terms down to appropriate forms.

\subsubsection{Adding it all together}
Adding all acquired terms together leads to a final expression for the residue
\begin{equation}
R_i = \frac{\partial z}{\partial u_i} G_{z_i} +  h_i \partial_{u_i} \log z'(u_i).
\end{equation}
By inserting the definition $G_i = \partial_{z_i} \log Q$ we can rewrite the residue to 
\begin{equation}
R_i = \partial_{u_i} \left( \log Q + h_i \log z'(u_i)\right).
\label{residue}
\end{equation}
This has the nice property of being a total derivative with respect to $u_i$, which is consistent with the Virasoro Ward identity. A second property is that this suggests a transformation rule for the probe sector of the correlator. It also, as was done in \cite{Anous:2019yku}, suggest a transformation rule for the probe sector of the correlator
\begin{equation}
\log Q' = \log Q +  h_i \log z'(u_i) \;\; \rightarrow \;\; Q'=\left(\frac{\partial z}{\partial u_i}\right)^{h_i} Q.
\label{transformationrule}
\end{equation}
This is the very familiar tensorial transformation rule of a primary operator of weight $h_Q$. Note that integrating \eqref{residue} for both light operator locations gives two integration constants. Demanding consistency between these integration constants (see appendix A for details) leads to
\begin{equation}
Q'=\left(\frac{\partial z}{\partial u_1}\right)^{h_1} \left(\frac{\partial z}{\partial u_2}\right)^{h_2} Q.
\end{equation}
This transformation property will be critical once we study the effect of residual global conformal symmetry below.

\subsection{Implications arising from the heavy sector accessory parameters}
Of course due to the genericity of the uniformizing coordinates at the light operators, the above analysis is just a derivation of the equivalence between the $TO$-OPE and the Virasoro Ward identity for primary operators. Something interesting occurs once we consider the accessory parameters associated with the heavy operators. We would like to perform a similar analysis around the heavy operators, notice though, that around these locations the uniformizing coordinate fails to be an analytic function, which could have serious consequences for the transformed stress tensor. In this short section I will show that the stress tensor maintains its meromorphic structure in the appropriate regimes.

Notice that in expression \eqref{uniformizedT} the second order poles have vanished at the heavy operator insertions, at the cost of introducing branch points at these locations on our background geometry. The asymptotics of the coordinate $u(z)$ in the vicinity of heavy operators is given by
\begin{equation}
u(z\rightarrow x_i) - u_{x_i}=(z-x_i)^{\alpha_i}(1+c_1 (z-x_i) + ...), \hspace{5mm} \alpha_i \equiv \sqrt{1-\frac{24H_i}{c}}.
\label{geometryasymptotics}
\end{equation}
Where we define $u_{x_i}$ as $u(u_{x_i}) \equiv x_i $. Now focus on the dominant term in the vicinity of $x_i$ 
\begin{equation}
T_Q(u\rightarrow u_{x_i}) = \left(\frac{\partial z}{\partial u}\right)^2 \frac{G_{x_i}}{z(u)-x_i}, \hspace{5mm} u(u_{x_i}) \equiv x_i.
\end{equation}
Now by writing out $G_{x_i}$ explicitly and taking the limit we find
\begin{align}
&\lim_{u\rightarrow u_{x_i}} \left(\frac{\partial z}{\partial u}\right)^2 \frac{\partial_{x_i}\log(Q)}{z(u)-x_i} \nonumber \\
&=\frac{\partial z}{\partial u}|_{u=u_{x_i}} \frac{\partial_{u_{x_i}}\log(Q)}{z(u)-x_i} \nonumber \\
&=\lim_{u\rightarrow u_{x_i}} \frac{1}{\alpha_i} (u(z)-u_{x_i})^{1/\alpha_i-1} \frac{\partial_{u_{x_i}}\log(Q)}{(u(z)-u_{x_i})^{1/\alpha_i}} \nonumber \\
&=\frac{\partial_{u_{x_i}} \frac{1}{\alpha_i} \log(Q)}{u(z)-u_{x_i}}
\end{align}
In the first step part of the Jacobian factor was absorbed into the derivative of $\log(Q)$ in the second step the explicit asymptotic form \eqref{geometryasymptotics} was inserted. From this final expression we can draw a couple of conclusions. First, $T_Q(u)$ remains a meromorphic function in the neighborhood of the image of a heavy operator. Secondly, the presence of a first order pole indicates that the probe correlator is charged under shifts with respect to the heavy operator locations. Finally we can conclude that there is no indication of the appearance of Jacobian factors with respect to the heavy operators, this will be the subject of the next subsection.



\subsection{Residual global conformal symmetry}
So far the uniformized stress tensor has been studied in two separate limits:
\begin{align}
& T_Q(u) = \left(\frac{dz}{du}\right)^2 \left( \sum_{i}^2 \frac{h_i}{(z(u)-z_i)^2} + \frac{G_{z_i}}{z(u)-z_i}+ \sum_{i}^n \frac{G_{x_i}}{z(u)-x_i}\right),\\
& \xrightarrow[]{u\rightarrow u_{z_i}}  \frac{h_i}{(u-u_{z_i})^2} + \frac{\partial_{u_{z_i}} \log Q'}{u-u_{z_i}},\\
& \xrightarrow[]{u\rightarrow u_{x_i}} \frac{\partial_{u_{x_i}}\log Q'}{u-u_{x_i}}.
\end{align}
The quantities in the numerators in these limits are given by
\begin{align}
& \partial_{u_{z_i}} \log Q' = \partial_{u_{z_i}} \log\left(\left(\frac{\partial z}{\partial u_{z_i}}\right)^{h_i} Q\right), \label{trafo1} \\
& \partial_{u_{x_i}} \log Q' = \partial_{u_{x_i}}\log \left(Q^{1/\alpha_i}\right).\label{trafo2}
\end{align}
The map $u(z)$ is not the unique conformal transformation that brings the stress tensor to uniformized form, but the remaining ones are all linked to $u(z)$ through means of global conformal transformations. This, combined with conformal invariance, gives us some power to constrain the dependence of the $u$-plane correlator on the light operator locations.

In order to study the effect of this freedom of global conformal frame conjoin the uniformizing geometry $u(z)$ with an $SL(2,\mathbb{C})$ transformation $w(u)$. Hence, the total transformation takes the form
\begin{equation}
w(u(z)), \hspace{5mm} w(u) = \frac{\alpha u + \beta}{\gamma u + \delta}, \hspace{5mm} \alpha \delta - \beta \gamma =1.
\end{equation}
It follows from the chain rule property of Schwarzian derivatives
\begin{equation}
S[w(u(z)),z] = S[u,z]+\left(\frac{\partial u}{\partial z}\right)^2 S[w,u]
\end{equation}
that $w(u(z))$ solves the ODE
\begin{equation}
T(w)=\frac{c}{12}S[z,w].
\end{equation}
One can trivially compute that $T_Q(w)$ can be expressed in terms of $T_Q(u)$ through
\begin{equation}
T_Q(w) = \left( \frac{\partial u}{\partial w}\right)^2 T_Q(u(w)).
\end{equation}
Which could have been guessed immediately from the group structure of global conformal transformations.

The interesting aspect of this transformation is that, as before, we can study the result of the transformation on the residues of the stress tensor. We can take over the calculations of the previous sections wholesale with one noticeable difference, the global conformal transformation $w(u)$ has no special properties at the locations of the heavy operators and can simply be Taylor expanded around the insertion points. Once again, the strategy is to take the Virasoro Ward identity seriously and associate the residues at the first order poles with the derivatives of (identity block contribution of) the conformal correlator in $w$-coordinates
\begin{equation}
Q''(w_{x_i},w_{z_i}) = \langle Q(w_{z_1})Q(w_{z_2}) O(w_{x_1}) ... O(w_{x_n})\rangle.
\end{equation}
The notation will be justified by following result which demonstrates that the new derivatives of the identity block only depend on the original light sector $Q(z_i,x_j)$. By extracting the residues of the stress tensor we find:
\begin{align}
& \partial_{w_{z_i}} \log Q''(w_{x_i},w_{z_i}) = \partial_{w_{z_i}} \log \left(\left(\frac{\partial u}{\partial w_{z_i}}\right)^h\ Q'(u(w_{z_i}),u(w_{x_i}))\right),\\
& \partial_{w_{x_i}} \log Q''(w_{x_i},w_{z_i}) = \partial_{w_{x_i}} \log \left(Q'(u(w_{z_i}),u(w_{x_i}))\right).
\end{align}
What this implies is quite particular, the dependence on the heavy insertions of the new correlator compared to the old one is trivial. Related by simple direct substitution. The light insertion dependence transforms in the familiar way, by picking a up a Jacobian factor, the derivative of the $SL(2,\mathbb{C})$-transformation. 

The striking conclusion is that the invariance under global conformal transformations after transforming to the uniformizing frame forces the light operator dependence to carry the full burden of invariance. This fixes the dependence of $Q'(u_{z_i},u_{x_i})$ to the form
\begin{equation}
Q'(u_{z_i},u_{x_i})= \frac{P(r_k(u_{z_i},u_{x_i}))}{(u_{z_1}-u_{z_2})^{2h}},
\label{ucorrelator}
\end{equation}
here $r_k(u_{z_i},u_{x_i})$ represent the various conformal cross-ratios that can be constructed out of the points $u_{z_i}, u_{x_i}$. In appendix A it will be demonstrated that cross-ratios that contain light operator locations are ruled out from appearing in $P(r_k)$, hence the two-point function factor in \eqref{ucorrelator} captures the full dependence of the $u$-plane correlator on the light operator insertions. If we pull this back to the original $z$-plane this results in the final expression for light sector dependence
\begin{equation}
Q(z_i,x_i)=\left(\frac{\partial u}{\partial z_1}\right)^h\left(\frac{\partial u}{\partial z_2}\right)^h\frac{1}{(u(z_1)-u(z_2))^{2h}},
\end{equation}
up to an overall multiplicative constant. This has the interpretation of a two-point function on a geometry given by the pullback $u(z)$. In the next section it will be shown that this matches the semi-classical Liouville correlator as long as the scaling weights of the vertex operators match the ones used above.

\section{Link to Liouville theory (...matter tells space how to curve)}
In the last section it was shown that the mixed identity block in the large-$c$ limit can be interpreted as a correlation function of light operators on top of a background geometry prepared by the light operators. This geometry was constructed through means of a differential equation that is closely related to the Liouville equation of motion. In this section it will be argued that this is not a coincidence, in fact it will be shown that the identity block contribution to semi-classical CFT correlators reproduces the same correlator within Liouville theory in $b\rightarrow 0$ limit. It should be emphasized that one should \textit{not} think of semi-classical Liouville theory as an example of a theory where the correlation functions are dominated by identity block exchanged, reasons being that Liouville theory has a continuous, i.e. ungapped, spectrum and no vacuum state. Instead the statement is that the Virasoro identity block in the large central charge limit approaches a semi-classical Liouville correlator. Let us set the conventions, which are taken near verbatim from \cite{Harlow:2011ny}, the action is given by
\begin{equation}
S_L=\frac{1}{4\pi} \int d^2 x \sqrt{\tilde{g}}\left(\partial_a\phi \partial_b \phi \tilde{g}^{ab} + Q \tilde{R}\phi + 4\pi \mu e^{2b\phi}\right).
\end{equation}
This action is coordinate invariant with respect to the background metric $\tilde{g}_{ab}$, as such it is conventional to simplify the action by rewriting it in terms of the flat metric 
\begin{equation}
S_L=\frac{1}{4\pi}\int_D d^2 x \left(\partial_a \phi \partial^a \phi + 4\pi\mu e^{2b\phi}\right) + \frac{Q}{\pi} \oint_{\partial D} \phi d\theta + 2Q^2\log(R).
\label{flatLiouville}
\end{equation}
The boundary integral term was added in order to `fix' the fact that the flat metric is incompatible with the Riemann sphere and the last term ensures finiteness of the action. Some details are in order, Q is given by
\begin{equation}
Q=b+\frac{1}{b},
\end{equation}
the central charge limit is given by 
\begin{equation}
c=1+6Q^2
\end{equation}
The semi-classical limit is obtained by the field redefinition $\phi \rightarrow \phi/2b$ and the subsequent limit $b\rightarrow 0$:
\begin{equation}
S_L=\frac{1}{16\pi b^2}\int_{D} d^2 x \left(\partial_a \phi \partial_a \phi + 16 \pi \mu b^2 e^{\phi}\right) +\frac{1}{2\pi b^2} \oint_{\partial D} \phi d\theta + \frac{2}{b^2} \log(R) + \mathcal{O}(b^2),
\label{scLiouville}
\end{equation}
here the combination $\mu b^2$ is assumed to be $\mathcal{O}(1)$ in the $b\rightarrow 0$ limit, this ensures that all terms explicitly written are $\mathcal{O}(1/b^2)$. The Liouville equation of motion is given by 
\begin{equation}
\partial\bar{\partial} \phi = 2\pi \mu b^2 e^{\phi}.
\end{equation}
At this point it is important to consider the spectrum of operators in Liouville theory. The action \eqref{scLiouville} is invariant under transformations of the form 
\begin{equation}
\phi(z,\bar{z})\rightarrow \phi(w(z),\bar{w}(\bar{z})) = \phi(z,\bar{z}) - Qb \log|\frac{\partial w}{\partial z}|^2.
\end{equation}
This logarithmic inhomogeneity suggests that it is the exponentials of the liouville field that transform as conformal primaries
\begin{equation}
\text{primary:} \hspace{3mm}  e^{\alpha \phi /b}, \hspace{5mm} \text{scaling dim.:} \hspace{3mm} ``h"=``\bar{h}"=\alpha Q.
\label{vertexscaling}
\end{equation}
The reason why these scaling weights are in quotation marks is that due to the fact that they pick up quantum corrections
\begin{equation}
L_0 e^{\alpha \phi/b} = \bar{L}_0 e^{\alpha \phi/b} = \alpha(Q-\alpha)e^{\alpha \phi/b}.
\end{equation}
This can be seen either by considering the regime of small field configurations where the exponential interaction term becomes negligible, in which case the correction term is due to the usual normal ordering procedure of free field vertex operators \cite{Harlow:2011ny}. Alternatively, it can be obtained by adding an improvement term to the classical Liouville stress tensor in order to allow for renormalization, fixing the improvement term by demanding that the resulting stress tensor has to satisfy the Virasoro algebra and reading off the new scaling dimensions of the vertex operator by computing the commutator of the new stress tensor with the vertex operator \cite{Seiberg:1990eb}.

Correlators are defined through a path integral over insertions of vertex operators
\begin{equation}
\langle e^{\alpha_1 \phi(z_1,\bar{z}_1)/b}...e^{\alpha_n \phi(z_n,\bar{z}_n)/b}\rangle = \int \mathcal{D}\phi e^{-S_L}e^{\alpha_1 \phi(z_1,\bar{z}_1)/b}...e^{\alpha_n \phi(z_n,\bar{z}_n)/b},
\end{equation}
the action $S_L$ is of the order $\mathcal{O}(b^2)$, hence we will distinguish between heavy operators where the term contributed to the action by the vertex operator is of the same order as the action and light operator where the vertex operator considers a term of order $\mathcal{O}(1)$. Explicitly:
\begin{align}
& \text{Heavy:} \; \eta_i = \lim_{b\rightarrow 0} \alpha_i b \propto \mathcal{O}(1),\\
& \text{Light:} \; \sigma_i = \lim_{b\rightarrow 0} \alpha_i/b \propto \mathcal{O}(1).
\end{align}
As a result, the correlation function of a string of heavy vertex operators shifts the equation of motion of the action to 
\begin{equation}
\partial \bar{\partial}\phi = 2\pi \mu b^2 e^{\phi} - 2\pi \sum_i \eta_i \delta^2(z-z_i).
\label{sourcedLiouville}
\end{equation}
This differential equation can be solved in the neighborhood of the delta function sources where the exponential term can be ignored, for that purpose apply the following identity for the delta function on the complex plane
\begin{equation}
\delta^2|z-z_i| = \frac{1}{\pi}\partial \frac{1}{\bar{z}-\bar{z}_i} = \frac{1}{\pi} \bar{\partial} \frac{1}{z-z_i},
\end{equation}
integrating twice gives respectively:
\begin{align}
& \partial \bar{\partial}\phi = -2\eta_i \partial \frac{1}{\bar{z}-\bar{z}_i} \;\; \rightarrow \;\; \phi = -2\eta_i \log(\bar{z}-\bar{z}_i) + f(z),\\
& \bar{\partial}\partial\phi = -2\eta_i \bar{\partial} \frac{1}{z-z_i} \;\; \rightarrow \;\; \phi = -2\eta_i \log(z-z_i) + g(\bar{z}).\\
\end{align}
Demanding consistency between the two and taking into account the order of the neglected terms gives us
\begin{equation}
\phi(z,\bar{z})= -4\eta_i \log|z-z_i| +\mathcal{O}(1).
\label{phiasympt}
\end{equation}
Let's consider the stress tensor, the leading order term in $b$ of the holomorphic stress tensor of the semi-classical action is given by
\begin{equation}
T(z)= -\frac{1}{4b^2} \left(\partial \phi\right)^2 + \frac{1}{2b}Q\partial^2 \phi.
\end{equation}
Plugging the above asymptotics \eqref{phiasympt} into the stress tensor we find 
\begin{equation}
T(z)= \left(\frac{\eta_i}{b}Q-\left(\frac{\eta_i}{b}\right)^2\right)\frac{1}{(z-z_i)^2} + \mathcal{O}\left(\frac{1}{z-z_i}\right),
\end{equation}
from this we can read off the scaling dimension of the heavy vertex operator
\begin{equation}
h=Q\alpha - \alpha^2,
\end{equation}
this demonstrates consistency with the scaling dimensions above and the Virasoro Ward identity. One might wonder how one can be sure that this stress tensor matches the one in the section above beyond just the second-order pole terms. The reason the accessory parameters match is because in the semi-classical CFT case these can be fixed through means of the Zamolodchikov monodromy method \cite{Hartman:2013mia}. In the case of the Liouville theory these can be fixed by demanding that the classical solution $\phi(z,\bar{z})$ results in a metric that is single-valued and positive definite everywhere. As was shown in \cite{Hulik:2016ifr}, when applying these two methods match exactly as long as one computes the vacuum block by monodromy method\footnote{This is also the reason why the method of the previous section implicitly specifies the identity block rather than any other  conformal block, the untransformed accessory parameters are assumed to be the same as those of the Liouville stress tensor.}.

\subsection{Liouville equation of motion}
The vacuum Liouville equation in the above conventions is given by
\begin{equation}
\partial\bar{\partial}\phi = 2l e^{\phi},
\end{equation}
where $l=\pi b^2 \mu$. By direct substitution it can easily be verified that this differential equation is solved by
\begin{equation}
e^{\phi}=\frac{1}{4l} \frac{|u'|^2}{(1-|u|^2)^2},
\label{Liouvillesolution}
\end{equation}
where $u(z)$ is any generic holomorphic function. We could include the sources, but we could also directly demand that $\phi$ has the right asymptotics \eqref{phiasympt} at the regular singular points, which is essentially all that the sources impose anyway. In order to accomplish this consider the following second order linear differential equation
\begin{equation}
\psi''(z) + b^2 T(z)\psi(z) = 0,
\label{LFuchs}
\end{equation}
this Fuchsian equation is of course well known in the uniformization literature, in addition it is also the null-vector decoupling equation that appears in the construction of conformal blocks \cite{Hartman:2013mia}. The reason for its occurence in uniformization is that once we define a new function $u(z)$ as the ratio of the independent solutions of the Fuchsian equation \eqref{LFuchs}, i.e. $u(z)=\psi_1(z)/\psi_2(z)$, then $u(z)$ will solve
\begin{equation}
\frac{1}{2}S[u,z]=b^2 T(z),
\label{LSchwarzian}
\end{equation}
where $S[u,z]$ is the Schwarzian derivative. This equation is familiar from the previous section, it is in fact exactly the inverse of the ODE \eqref{inverseschwarzianode}, as the stress tensor $T(z)$ matches the undeformed heavy stress tensor expectation value. The remarkable property of $u(z)$ is that once it is inserted into the Liouville solution \eqref{Liouvillesolution} $\phi(z)$ will automatically satisfy the right conditions \eqref{phiasympt} near the operator insertions. This can easily be checked by taking
\begin{equation}
T(z\rightarrow z_i) = \frac{\rho}{(z-z_i)^2},
\end{equation}
in which the case the Fuchsian equation is easily solved. Subsequently constructing $u(z)$ by taking the ratio of solutions and inserting this solution into \eqref{Liouvillesolution}. Take the limit $z\rightarrow z_i$ and $\phi(z)$ will reproduce the asymptotics \eqref{phiasympt}, given that we identify $\rho=\eta/b^2-\eta^2/b^2$, i.e. $h$ in the $b\rightarrow 0$ limit.

\subsection{Transformation properties}
It will be useful to establish some transformation properties of the above objects under conformal transformations. As we will see this will be enough to fix the semi-classical correlator. First, as is well known, the stress tensor $T(z)$ transforms as a coadjoint vector of the Virasoro algebra. In particular the combination $b^2 T(z)$ in the $b \rightarrow 0$ regime transforms under $v(z)$ as
\begin{equation}
b^2 T(z) \rightarrow b^2 T(v) = \left(\frac{\partial z}{\partial v}\right)^2 b^2 T(z(v)) + \frac{1}{2}S[z,v].
\end{equation}
It is straightforward to check that if the above transformation rule is substituted in the ODE \eqref{LFuchs} that the resulting equation is solved by
\begin{equation}
\psi(z) \rightarrow \tilde{\psi}_v(z) = \frac{1}{\sqrt{v'(z)}} \psi(v(z)).
\end{equation}
This leads to the interesting conclusion that the ratio of independent solutions $u(z)=\psi_1(z)/\psi_2(z)$ transforms as a scalar under conformal transformations
\begin{equation}
u(z) \rightarrow u(v(z)),
\end{equation}
alternatively this could have been checked by transforming the right-hand side of \eqref{LSchwarzian} and taking into account the chain rule property of the Schwarzian derivative. Notice also that, by the chain rule, this results in consistent transformation rules on both the left- and right-hand side of \eqref{Liouvillesolution}\footnote{Within this context $\exp(\phi)$ is just the exponent of a classical solution, not a quantum operator, hence the `scaling weight' of $\exp(\phi)$ doesn't contain any quantum corrections.}.

One last thing to mention is that we are free to choose a different linear basis to refer to as $\psi_1(z)$ or $\psi_2(z)$, this choice does affect $u(z)$. If we transform to a different basis through
\begin{equation}
\psi_i(z) \rightarrow M_{ij}\psi_j(z) \;\;\;\; M_{ij}=
\begin{pmatrix}
a&b\\
c&d\\
\end{pmatrix}
\in SL(2,\mathbb{C}),
\end{equation}
this will affect $u(z)$ through
\begin{equation}
u(z) \rightarrow \frac{a u(z) +b}{cu(z)+d}.
\end{equation}
It is helpful to compare and contrast the difference between applying a basis transformation and a global conformal transformation. Both are $SL(2,\mathbb{C})$ transformations represented by M\"obius transformations $M(z)$ but the distinction is that a global conformal transformation affects $u(z)$ through $u\circ M$ whereas the effect of a basis transformation is given by $M\circ u$, these operations of course do not commute.

One final remark, the general solution to \eqref{LFuchs} is given by a generic combination of independent linear solutions. By the above considerations this implies that the general solution $u_g$ to \eqref{LSchwarzian} is given by an $SL(2,\mathbb{C})$ orbit of some representative specific solution $u_s$:
\begin{equation}
u_g(z) = \frac{a u_s(z) +b}{c u_s(z) +d} \;\;\;\; ad-bc=1,
\end{equation}
this will be relevant as in the saddle-point method it is relevant that one sums over \textit{all} dominant saddles, hence resulting in an integral over the moduli space of $SL(2,\mathbb{C})$.

\subsection{Semi-classical correlators}
We will deal with Liouville correlators of the mixed heavy-light kind in the semi-classical limit, a regime that is particularly well-suited for saddle point methods. The action in the $b\rightarrow 0$ limit was given in the previous section
\begin{equation}
S_L=\frac{1}{16\pi b^2}\int_{D} d^2 x \left(\partial_a \phi \partial_a \phi + 16 \pi \mu b^2 e^{\phi}\right) +\frac{1}{2\pi b^2} \oint_{\partial D} \phi d\theta + \frac{2}{b^2} \log(R) + \mathcal{O}(b^2).
\end{equation}
The correlators of interest are those that contain two light operators and an arbitrary (though \textit{far} fewer than $\mathcal{O}(c)$) $n$ number of heavy operators
\begin{equation}
\langle e^{\eta_1 \phi(x_1)/b^2}...e^{\eta_n \phi(x_n)/b^2} e^{\sigma_1 \phi(z_1)}e^{\sigma_2 \phi(z_2)}\rangle = \int \mathcal{D}\phi\, e^{S_L[\phi] - \frac{1}{b^2}\left(\sum_{i=1}^n \eta_i \phi(x_i)\right) - \sigma_1 \phi(z_1) - \sigma_2 \phi(z_2)}. 
\label{pathintegral}
\end{equation}
To leading order in $b$ the saddle-point configuration of $\phi_{s}$ is independent of $z_i$ and given by the Liouville equation
\begin{equation}
\partial \bar{\partial}\phi_s = 2\pi \mu b^2 e^{\phi_s} - 2\pi \sum_{i=1}^n \delta^2(z-x_i),
\end{equation}
As discussed above, the general solution is given by
\begin{equation}
e^{\phi_s} = \frac{\frac{|u'(z)|^2}{|c u(z) + d|^2}}{(1-|\frac{au(z)+b}{cu(z)+d}|^2)^2}, \;\;\; ad-bc=1
\label{generalsolution}
\end{equation}
where $u(z)$ is a particular solution to
\begin{equation}
\frac{1}{2}S[u,z] = b^2 T(z).
\end{equation} 
If we want to resolve the path integral \eqref{pathintegral}, which means summing over all dominant saddle points. The saddle points are given by the set of particular solutions parametrized by $a, b, c$ and $d$, hence we will have to integrate over these moduli. One important thing to note, the subgroup $SU(1,1)$ of $SL(2,\mathbb{C})$ leaves the right hand-side of \eqref{generalsolution} invariant and hence the saddle point invariant, therefore the integral is over the moduli of $SL(2,\mathbb{C})/SU(1,1)$. The exact representation of this subgroup of $SL(2,\mathbb{C})$ does not matter, this is because $SL(2,\mathbb{C})$ acts transitively on matrix subgroups as a result of which any isomorphism can be incorporated by a linear change of integration variables with trivial Jacobian density. 

Substituting the saddle points back into the correlation function \eqref{pathintegral} results in
\begin{align}
& \langle e^{\eta_1 \phi(x_1)/d^db^2}...e^{\eta_n \phi(x_n)/b^2} e^{\sigma_1 \phi(z_1)}e^{\sigma_2 \phi(z_2)}\rangle = \int d^2a d^2b d^2c d^2 d\, \delta(ad-bc-1) \nonumber\\
& \times\left(\frac{\frac{|u'(z_1)|^2}{|cu(z_1)+d|^2}}{\left(1-|\frac{au(z_1)+b}{cu(z_1)+d}|^2\right)^2}\right)^{\sigma_1}\times\left(\frac{\frac{|u'(z_2)|^2}{|cu(z_2)+d|^2}}{\left(1-|\frac{au(z_2)+b}{cu(z_2)+d}|^2\right)^2}\right)^{\sigma_2}.
\label{integral}
\end{align}
What this means is that we are integrating over a subset of all M\"obius transformations of $u(z_i)$. We can exploit this fact to shift the `starting point' of the sum over the $SL(2,\mathbb{R})$ orbit to a more convenient location. Take the transformation
\begin{equation}
\tilde{w}=\frac{u(z)-u(z_1)}{u(z)-u(z_2)},
\end{equation}
this has the effect of reducing the integral \eqref{integral} to
\begin{align}
& \langle e^{\eta_1 \phi(x_1)/d^db^2}...e^{\eta_n \phi(x_n)/b^2} e^{\sigma_1 \phi(z_1)}e^{\sigma_2 \phi(z_2)}\rangle = \nonumber \\ 
& \frac{|u'(z_1)|^{2\sigma_1}|u'(z_2)|^{2\sigma_2}}{|u(z_1)-u(z_2)|^{4\sigma_1}} \int d^2a d^2b d^2c d^2 d\, \delta(ad-bc-1) \left(|d|^2-|b|^2\right)^{-2\sigma_1}\times\left(|c|^2-|a|^2\right)^{-2\sigma_2}.
\label{splitoffintegral}
\end{align}
Hence the entire integral seems to reduce to nothing more than a multiplicative constant. This is not quite true, as demonstrated in appendix B the integral vanishes identically almost everywhere, an exception being when $\sigma_1=\sigma_2\equiv \sigma$ hence this factor will be interpreted as providing the usual CFT constraint that two-point functions of primary operators vanish except when their scaling dimensions match. This takes care of the asymmetry in $\sigma_1$ and $\sigma_2$ in the factor outside of the integral. The complete $w_i$ dependence of the correlator is therefore given by
\begin{equation}
\langle e^{\eta_1 \phi(x_1)/d^db^2}...e^{\eta_n \phi(x_n)/b^2} e^{\sigma_1 \phi(z_1)}e^{\sigma_2 \phi(z_2)}\rangle = \frac{|u'(z_1)|^{2\sigma}|u'(z_2)|^{2\sigma}}{|u(z_1)-u(z_2)|^{4\sigma}}.
\end{equation}
This expression factorizes into the product of an holomorphic and anti-holomorphic. In the $b\rightarrow 0$ limit, the holomorphic factor is in perfect agreement with the light sector of the large-$c$ identity block computation from the previous section. This demonstrates that the semi-classical identity block in a generic CFT reproduces the same correlator in semi-classical Liouville theory.

\section{HHLL done triply well}
One of the most interesting special cases is when the heavy sector consists of just two operators. If one of these operators is located at the origin and the other at infinity then the correlation function studied so far is of the HHLL type and in particular has the quantum interpretation of the expectation value of the product of two light operators on an energy eigenstate of energy $H$:
\begin{equation}
Q(z_1,z_2) = \langle H|Q(z_1)Q(z_2)|H\rangle.
\end{equation}
This has been used in the past to study the realization of the eigenstate thermalization hypothesis (ETH) in semi-classical CFTs \cite{Fitzpatrick:2015zha,Anous:2019yku}, and the more extensive correlator HHLLLL which has been used to study the quantum butterfly effect \cite{Anous:2019yku}.

In this particular case the heavy sector stress tensor is given by 
\begin{equation}
\frac{6}{c}T(z)  = \frac{ 6H_1/c}{z^2},
\end{equation}
the resulting Schwarzian equation can be solved by a power law solution
\begin{equation}
u(z)=z^{\alpha_i},
\end{equation}
where $\alpha_i$ is defined through
\begin{equation}
\alpha_i = \sqrt{1-24H_i/c}.
\end{equation}
One the one hand this is the simplest possible case, since all differential equations above can be solved in terms of simple elementary functions. On the other hand, this set-up is subject to a surprising level of nuance at the geometric level. For that we first turn to a simple well-known constraint \cite{Seiberg:1990eb} due to the Gauss-Bonnet theorem. Which on surfaces without boundary reads as:
\begin{equation}
\int d^2 x \, \sqrt{g} R = 4\pi (2-2g),
\label{GaussBonnet}
\end{equation}
where $g$ is the genus of the surface. For two-dimensional metrics in the conformal gauge the Ricci scalar is given by:
\begin{equation}
R = -e^{-\phi}\Delta \phi 
\end{equation}
Substituting the sourced Liouville equation \eqref{sourcedLiouville} into the Ricci scalar and inserting the result into \eqref{GaussBonnet} gives us
\begin{equation}
2g-2 + \sum_i^n \eta_i = b^2 \mu A,
\end{equation}
where $A$ represents the total area of the surface. Demanding that the total area of our surface is positive leads to the famous inequality:
\begin{equation}
2g-2 + \sum_i^n \eta_i > 0.
\end{equation}
This is a geometrical statement which can be rephrased into CFT language expressing the heavy vertex operator exponent in terms of a scaling dimension
\begin{equation}
H_i=\frac{1}{b^2}(\eta_i - \eta_i^2),
\end{equation}
inverting this and substituting it in the Gauss-Bonnet bound yields
\begin{equation}
2g+\frac{1}{2}n-2-\frac{1}{2}\sum_i^n \alpha_i >0.
\end{equation}
There are a couple of things to note here. First, $\alpha_i$ becomes imaginary when $H>c/24$, this is the Seiberg bound and reflects the fact that there are no local operators in Liouville theory with scaling weights over $c/24$. Our generic CFT is not subject to such spectral constraints. Secondly, it is impossible to satisfy the Gauss-Bonnet bound with only two operators inserted that do satisfy the Seiberg bound inserted on the Riemann sphere. This is not a problem, it simply reflects the uniformization theorem; not all Riemann surfaces can be pulled back to a quotient of the upper half-plane (equivalently, the Poincar\'e disk) some can be pulled back to a quotient of the Riemann sphere or the complex plane. 

The situation with two local (or microscopic \cite{Seiberg:1990eb}) is one of the cases that can be mapped to a quotient of the sphere. The presence of the punctures does not add enough curvature to overcome the natural intrinsic curvature of the sphere. Hence unlike \eqref{Liouvillesolution}, the natural metric for the space on which the light operators live is given by
\begin{equation}
e^{\phi}=\alpha_1^2 \frac{|z|^{2(\alpha_1-1)}}{(1+|z|^{2\alpha_1})^2},
\end{equation}
notice that for convenience $l$ has been set to $\frac{1}{4}$ this will be true for the whole section. This metric is conventionally obtained by adding a Lagrange multiplier to the Liouville path integral that fixes the area of the Riemann surface and then regulating an integral over all areas \cite{Seiberg:1990eb, Zamolodchikov:1982vx}. Notice the plus sign in the denominator, this metric becomes the familiar metric of the sphere in the limit $H\rightarrow 0$. For generic values of $\alpha_1$ this metric describes a sphere where a wedge with angle $2\pi(1-\alpha_1)$ running from the north to the south pole has been cut out and the resulting edges are sewn together. The result is a sphere with two antipodal dimples.\\

\noindent Now let's consider the case of two heavy operators that do not satisfy the Seiberg bound. Within our original CFT there was no Liouville action subject to regularity conditions, all we did was arrive at a differential equation that functions as an auxiliary device towards solving the Liouville equation with sources. The link to Liouville equation gives us a beautiful geometric picture of semi-classical CFT and a direct link to gravity in AdS$_3$, but the differential equation remains perfectly well-defined once we relinguish the Seiberg bound. This case has been studied quite extensively and was shown to correspond to stationary black hole physics in AdS$_3$ \cite{Fitzpatrick:2015zha,Anous:2019yku}. 

\begin{figure}
	\centering
		\includegraphics[scale=0.15]{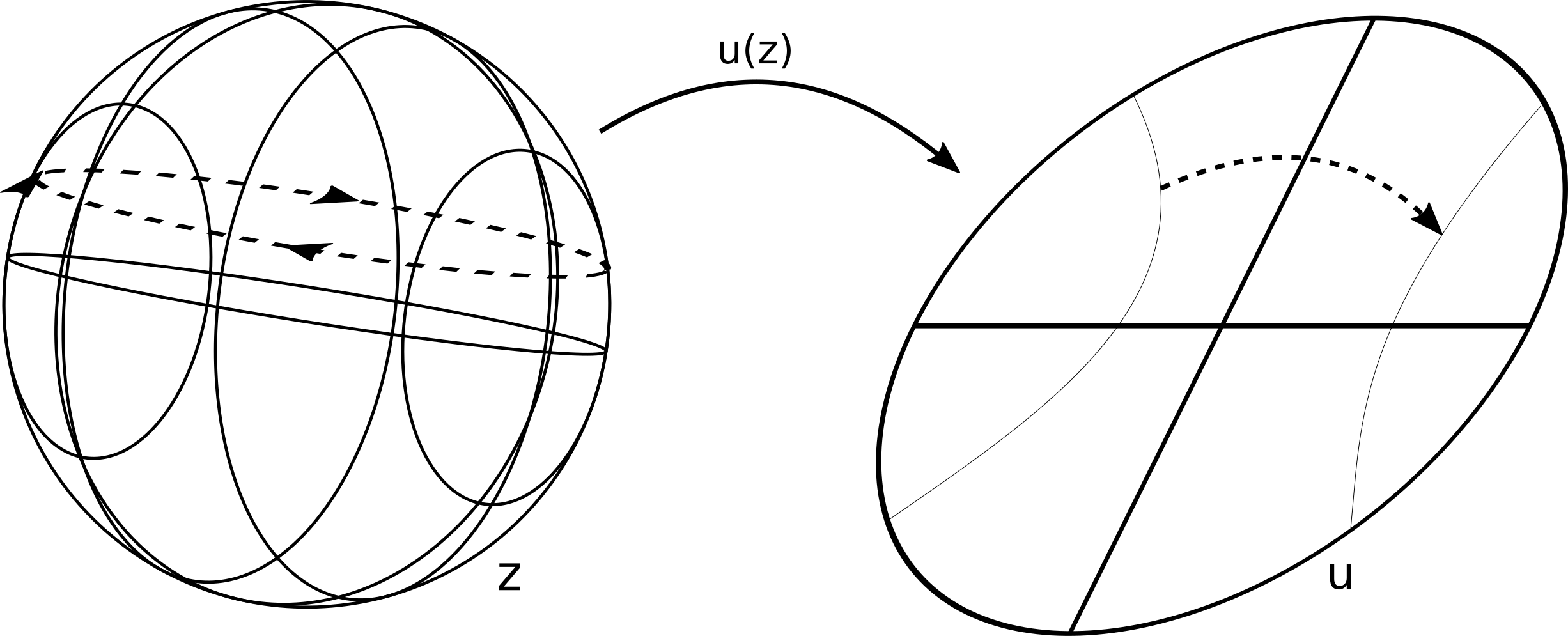}
	\caption{The map $u(z)$ functions as a map from a coordinate on the Riemann sphere to a coordinate on the Poincar\'e disk. Closed circles on the Riemann sphere map to isometry transformations on the Poincar\'e disk. Therefore, demanding that the metric on the Poincar\'e disk is single-valued as a function of $z$ naturally implies that it describes a space given by a quotient of the Poincar\'e disk by the isometry transformation. A basis for $u$ has been chosen that exemplifies the link to the BTZ geometry.}
	\label{identificationpicture}
\end{figure}

This can be corroborated geometrically through the machinery of uniformization through Fuchsian discrete groups, something which is some sense analogous to what was done in the example above where we cut a wedge out of the Riemann sphere. Consider now the metric in the case of hyperbolic punctures
\begin{equation}
e^{\phi}=\alpha_1^2 \frac{|z|^{2(\alpha_1-1)}}{(1-|z|^{2\alpha_1})^2},
\end{equation}
notice the minus sign in the denominator and the fact that $\alpha_1$ is now assumed to be purely imaginary. At first sight the Weyl factor appears to be multi-valued on the Riemann sphere. This is not a problem as the monodromy one picks up when circumnavigating the origin is of the form
\begin{equation}
z^\alpha \rightarrow e^{2\pi i\alpha}z^{\alpha} = e^{-2\pi \sqrt{24 H/c -1}}z^{\alpha} \equiv \frac{z^{\alpha}}{\lambda}, \;\;\; H>\frac{c}{24},
\label{generator}
\end{equation}
this is a fractional linear transformation of loxodromic class and in particular it falls within $SU(1,1)$ subgroup of $SL(2,\mathbb{C})$ that leaves the unit disk and, in this case, the metric invariant. This forms an equivalence relation on the Poincar\'e disk which reduces the space to a quotient $\mathbb{H}/\Gamma$, where $\mathbb{H}$ is the Poincar\'e disk and $\Gamma$ is the discrete Fuchsian group freely generated by the single hyperbolic generator \eqref{generator}, see figure \ref{identificationpicture}.

\begin{figure}
	\centering
		\includegraphics[scale=0.18]{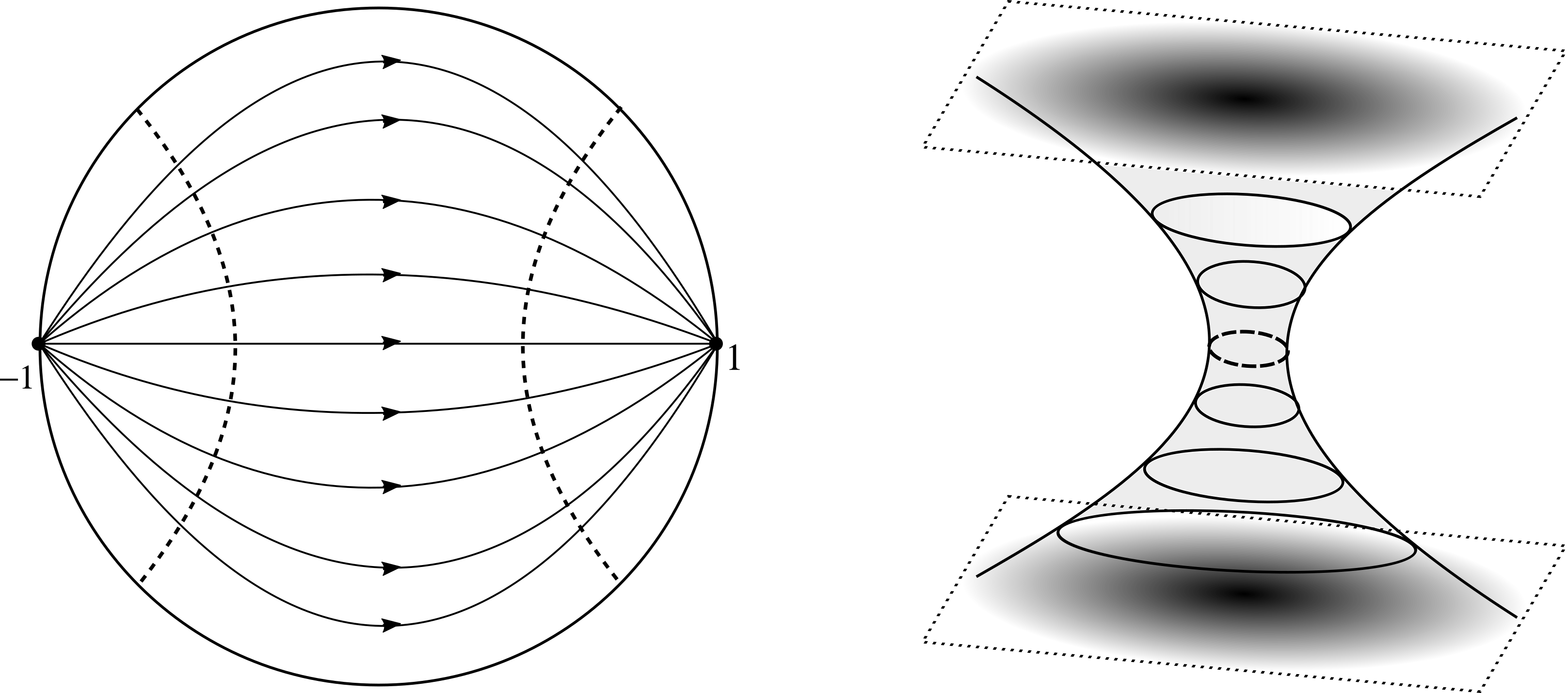}
	\caption{A reproduction of a figure in \cite{Krasnov:2000zq}. The image on the left represents the fundamental domain of $\mathbb{H}/\Gamma$. The flow lines connecting the points 1, -1 are the curves that are mapped onto itself by the hyperbolic generator. The dashed lines cut all the flow exactly once and perpendicularly, they are mapped onto each other by the hyperbolic generator. It is these dashed lines that are identified. The right image shows an embedding of the left fundamental domain in three-dimensional space. Due to the metric of $\mathbb{H}$ this takes the form of a hyperboloid. The dashed line represents the non-contractible cycle of minimal length, its length $A$ is given by $A=\log{\lambda}$.}
	\label{FundamentalDomain}
\end{figure}

The relationship between the fundamental domain and the BTZ black hole is obscured because the second fixed point of the hyperbolic generator is located on a universal cover of $\mathbb{H}$. This can be fixed by conjugating the Poincar\'e coordinate $u$ by a M\"obius transformation such that the fixed points of the transformed generator are given by 1 and -1. Constructing the appropriate generator is simple as it is determined entirely by its trace and fixed points, the result is
\begin{equation}
\text{Fuchsian generator:} \;\;\;\;  u \rightarrow \frac{\frac{1}{2}\gamma u + \sqrt{\gamma^2/4-1}}{u \sqrt{\gamma^2/4-1}+\frac{1}{2}\gamma}, \;\;\; \gamma \equiv \lambda + \frac{1}{\lambda}.
\end{equation}
This generator makes it particularly easy to connect the fundamental domain of $\mathbb{H}/\Gamma$ with the BTZ black hole as the approach from this point on is entirely textbook material. Hyperbolic transformations of the Poincar\'e disk have the property of moving points a fixed distance along flow lines that connect the two fixed points, see figure \ref{FundamentalDomain}. We can now construct a fundamental domain of $\mathbb{H}/\Gamma$ by constructing a line that crosses every flow line exactly once perpendicularly. Construct image of this line under the hyperbolic generator and identify the points on these lines with their images under the generator. This results in non-compact strip on $\mathbb{H}$ which plays the role of the fundamental domain of $\mathbb{H}/\Gamma$.

This space can be naturally embedded in three-dimensional space that explicitly shows the periodic identification. This space is not simply connected and in fact is the usual spatial slice of zero extrinsic curvature of the BTZ black hole. This space has a non-contractible cycle that can be contracted down to a minimum value that is very simple to compute. One of the flow lines of the generator is the geodesic connecting -1 to 1. Every two points connected by the generator are separated by the same geodesic distance on $\mathbb{H}$, so as an example we can take the origin which is mapped to $\sqrt{1-4/\gamma^2}$, this results in a geodesic length A of size
\begin{equation}
A=\int_0^{\sqrt{1-4/\gamma^2}} \frac{dr}{1-r^2} = \log{\lambda} = 2\pi \sqrt{24H/c-1}.
\end{equation}
This length can be recognized as the horizon area of a BTZ black hole dual to a primary state of scaling weight $H>c/24$, hence the minimal length of the family of non-contractible cycles on the geometry measures the size of the event horizon. The Hawking temperature of this non-rotating black hole is given by
\begin{equation}
T_{BTZ} = \frac{A}{4\pi^2} = \frac{1}{2\pi}\sqrt{24H/c-1}.
\end{equation}
This is exactly the temperature experienced by the light operators \cite{Fitzpatrick:2015zha,Anous:2019yku,Banerjee:2018tut}

\begin{figure}
	\centering
		\includegraphics[scale=0.08]{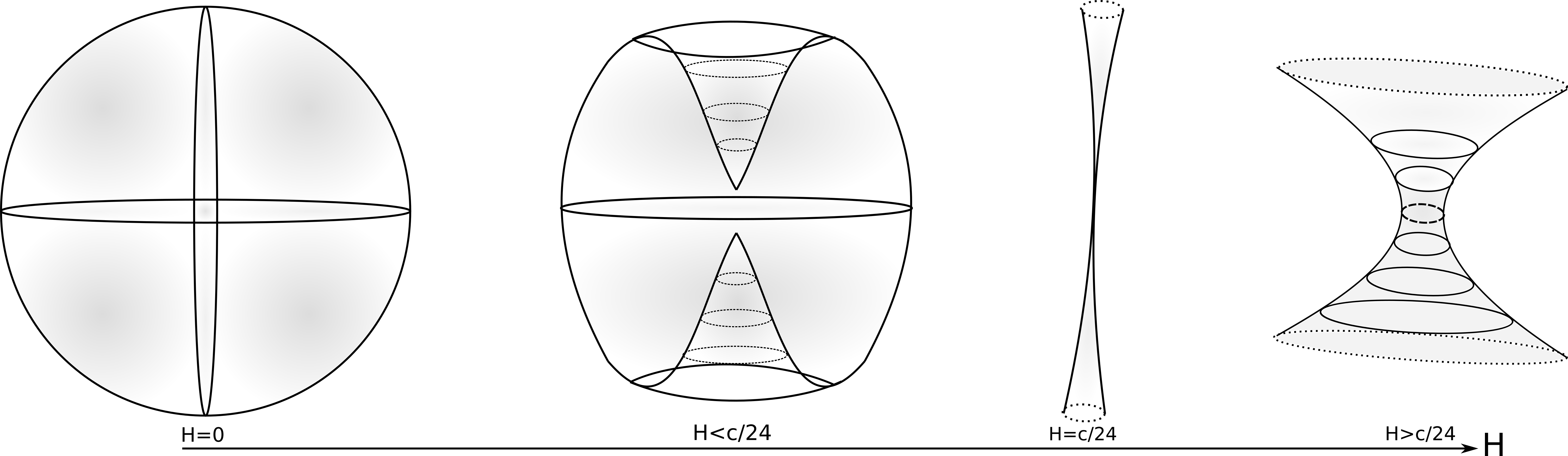}
	\caption{A set of examples of effective geometries of the primary state $|H\rangle$ as the scaling weight $H$ is varied. Notice the transition in topology at $H=c/24$.}
	\label{fig:Geometries}
\end{figure}

There has been some discussion lately from the CFT perspective as to this transition. This transition appears to cross from an ergodic non-thermalizing phase to one where the state does act as a thermal state. Keeping track of the geometries as one varies the scaling weight $H$ we see a violent change in topology of the effective background as $H$ crosses the thermal transition point at $H=c/24$, see figure \ref{fig:Geometries}. Therefore these operators play the role of topology changing operators familiar from matrix theories or orbifold models. Within the CFT, primary states with scaling weights $H<c/24$ act as ergodic states with respect to light operators, while states with $H>c/24$ act as thermal backgrounds for probe operators \cite{Fitzpatrick:2015zha,Anous:2019yku}. It is telling to see that at the transition point the effective geometry changes topology to one with a non-contractible cycle with a length proportional to the temperature.

I will close this section with one last comment on conical defects; since conical defect on the Poincar\'e disk can be constructed, it is interesting to see that these don't manifest as effective geometries for the light particles. The reason essentially being that the Riemann sphere can't be forced to `fit' inside the Poincar\'e disk. This is particularly strange once we discuss the holographic interpretation in the next section. It is tantalizing to notice though that if instead of the dimpled sphere the light operators where to be placed onto the Poincar\'e disk with a conical defect by hand, if one again integrates over the moduli space of $SL(2,\mathbb{R})$ the result would still match the HHLL correlator. This is essentially due to the fact that the required integrals over the group manifold are related by a trivial change of integration variables.




\FloatBarrier

\section{Holographic implications and future directions}
The result above indicates that the Virasoro identity block is given by a semi-classical Liouville correlator. This possesses a clear interpretation in CFTs with potential semi-classical bulk duals. These models possess a gap in the operator spectrum between the identity operator and the first non-trivial primary operator. Hence operators can be spaced out in such a way that the dominant contribution to the correlator is the identity block. There is a natural holographic interpretation, the vacuum block contains the exchanges of all products and derivatives of the stress tensor. Within the usual holographic dictionary the stress tensor is dual to the graviton, hence in the bulk the vacuum block sums up all ladder diagrams of internal graviton lines. Hence, this regime has a gravitational dual description as the regime where gravity is the dominant force in exchanges. 

The emergence of Liouville theory in the regime of interest is salient as Liouville theory is well-known to be equivalent to asymptotically AdS$_3$ Einstein gravity at the classical level. The original argument was presented in \cite{Coussaert:1995zp} and reviewed in \cite{Carlip:2005zn}. It can be roughly summarized as follows, impose the Brown-Henneaux boundary conditons \cite{Brown:1986nw} on aymptotically AdS$_3$ Einstein-Hilbert action.  After rewriting the Einstein-Hilbert action into Chern-Simons theory the boundary conditions manifest themselves as asymptotic boundary conditions on the gauge fields. Part of these boundary conditions can be solved to reduce the Chern-Simons theory to a WZW-theory. Finally, the remaining boundary conditions implement the Hamiltonian reduction from WZW to Liouville theory. Once the dust settles this informs us that at the classical level Liouville theory captures the physics of asymptotically AdS$_3$ physics with Brown-Henneaux boundary conditions.

Secondly, Liouville theory has a more direct realization in AdS$_3$ gravity through the ADM formalism. Take a gauge where AdS$_3$ is foliated by time slices in such a way that there is at least one slice with zero extrinsic curvature. The Hamiltonian initial slice constraint takes the form of the Liouville equation \cite{carlip}. A solution to this constraint combined with the zero extrinsic curvature assumption is enough initial data to construct an entire Lorentzian AdS$_3$ space that solves the Einstein equations. It is tempting to identify the geometry generated by the heavy operators as the geometry of this particular time slice. In this case one would interpret the light sector of the identity block as a probe limit where the light bulk particles move through the heavy background geometry without backreaction. One way to prove this semi-classical picture would be to interpret the ADM formalism as a Schr\"odinger evolution of $\langle H...H|QQ|H...H\rangle$ and compare this with the Heisenberg time-evolution of \cite{Banerjee:2018tut}. It should also be mentioned that the exercise involving the Fuchsian uniformization of the double marked sphere can be extended to larger collection of marked points. This can be done by considering Fuchsian groups generated by greater numbers of loxodromic generators \cite{Krasnov:2000zq,Brill:1998pr}. It is a difficult problem to obtain quantitative results for these kinds of geometries, but basic questions such as the future topology after Lorentzian time-evolution can be tackled qualitatively.

One might be concerned as to how such local information in the bulk is encoded within an exercise involving local operators in the CFT. At first sight this seems to clash with the usual AdS/CFT intuition that local bulk operators correspond to highly non-local CFT operators. There are two reasons for this. First, the large $c$ limit causes the bulk physics to reduce to classical gravity, hence eliminating any quantum gravitational subtleties. Secondly the relationship presented in this paper is between Euclidean CFT and AdS$_3$. The relationship between Euclidean AdS/CFT and Lorentzian AdS/CFT, while there, is highly non-trivial. (Since a couple of years, H. Verlinde has been referring to this set-up as CFT/AdS in order to explicitly distinguish it from AdS/CFT).


Finally, this discussion has focused on the case of two light operators. This was due to a practical reason, two-point function could be fixed by the residual global conformal invariance. It seems natural though to conjecture that the general principle should hold for greater products of light operators as long as the sum of the scaling dimensions remains far smaller than the central charge.

\section*{Acknowledgements}
The author of this paper has benefited greatly from various discussions over the course of this project, some of these people include: T. Anous, O. Hul\'ik, K. Papadodimas, J. Raeymaekers, J. Sonner, E. Verlinde. This work was supported  by the Grant Agency of the Czech Republic under the grant EXPRO 20-25775X.


\appendix

\section{Ruling out light operator cross-ratios}
In order to rule out the appearance of cross-ratios depending on light operator insertions in \eqref{ucorrelator} I will use the following simple strategy. We take the solution for $Q'$ and substitute it back into the original $z$-plane correlator through means of \eqref{trafo1} and \eqref{trafo2}. Integrating both sides of those equations leads to a set of integration constants (in reality are functions of all variables except the respective one that is integrated over). The difference between \eqref{trafo1} and \eqref{trafo2} reflects an asymmetry in the dependence of the light sector on the heavy insertions and light insertions. Demanding consistency between this asymmetry and the integration constants leads to a constraint that proves the desired result. 

First integrate and subsequently take the exponent of \eqref{trafo1}:
\begin{equation}
Q'(u_{z_i},u_{x_i})=\left(\frac{\partial z}{\partial u}|_{u_{z_i}}\right)^h Q(z(u_{x_i}),z(u_{z_i})) M_i(u_{z_{j\neq i}},u_{x_k}),
\end{equation}
here $M_i(u_{z_{j\neq i}},u_{x_i})$ are arbitrary functions of potentially all points except for $u_{z_i}$. Performing a similar process for \eqref{trafo2} results in:
\begin{equation}
Q'(u_{z_i},u_{x_i})= Q(u_{z_i},u_{x_i})^{1/\alpha_i}N_i(u_{z_{k}},u_{x_{j\neq i}}),
\end{equation}
the same goes for $N_i$ as does for $M_i$, except it is now a heavy insertion point that is excluded.

Both these expressions above can be solved for $Q$, substituting our knowledge of $Q'$ through \eqref{ucorrelator} results in:
\begin{align}
& Q=\frac{P(r_k)}{\left(\frac{\partial z}{\partial u}|_{u_{z_i}}\right)^h (u_{z_1}-u_{z_2})^{2h} M_i}, \label{Qconstraint1} \\
& Q=\left( \frac{P(r_k)}{(u_{z_1}-u_{z_2})^{2h} N_i}\right)^{\alpha_i} \label{Qconstraint2}. 
\end{align}
First things first, left-hand side of \eqref{Qconstraint1} is independent of the index $i$, i.e. the right-hand sides have to match for each value of $i$. Equating the two gives us some simple unsurprising information about $M_i$:
\begin{equation}
\left(\frac{\partial z}{\partial u}|_{u_{z_1}}\right)^h M_1 = \left(\frac{\partial z}{\partial u}|_{u_{z_2}}\right)^h M_2.
\end{equation}
Since $M_i$ is independent of $u_{z_i}$ this gives us a simpler expression
\begin{equation}
M_i=\left( \frac{\partial z}{\partial u}|_{u_{z_{j\neq i}}}\right)m(u_{x_k}),
\end{equation}
where $m(u_{x_k})$ is a function that can only depend on the heavy operator locations, i.e the light sector dependence of $M_i$ can be explicitly factored out.

Now we can move on to constricting $P(r_k)$, in order to accomplish this equate one of the expressions \eqref{Qconstraint1} with one of \eqref{Qconstraint2} and solve for $P(r_k)$
\begin{equation}
P(r_k)= (u_{z_1}-u_{z_2})^{2h} \left( \frac{N_i^{\alpha_i}}{\left( \frac{\partial z}{\partial u}|_{u_{z_1}}\right)^h \left(\frac{\partial z}{\partial u}|_{u_{z_2}}\right)^h m(u_{x_k})}\right)^{\frac{1}{1-\alpha_i}}.
\end{equation}
The argument ruling out cross-ratios in $P$ that mix light and heavy operator locations can be completed through proof-by-contradiction. Imagine that the right-hand side depends on both a point $u_{x_l}$ and $u_{z_j}$, since the only remaining free function that is allowed to depend on $u_{z_j}$ is $N_i$ it is these function that carry the burden of ensuring that the light sector dependence on both sides of the equations match. Notice though that when $i=l$ the function $N_l$ is not allowed to depend on the heavy point involved in the cross-ratio. Since the left-hand side does depend on $u_{x_l}$ by assumption, this implies that it is impossible for the right-hand side to reproduce the exact same left-hand side for every index $i$. Notice however that this argument cannot be used to rule out that $P(r_k)$ consists exclusively of cross-ratios of the heavy insertion locations.

\section{Integrating over the $SL(2,\mathbb{C})$ moduli space}
In the body of the text during the evaluation of the mixed heavy-light correlator in Liouville theory an integral over the moduli space of $SL(2,\mathbb{C})$ was encountered. It was claimed that while this integral does not contain any dependence on operator positions it does play the role of providing a constraint that demands both light operators have the same scaling dimension. This appendix will elaborate on that point. The integral in \eqref{splitoffintegral} is the one under consideration, i.e.
\begin{equation}
I(\sigma_1,\sigma_2)=\int d^2a d^2b d^2c d^2 d\, \delta(ad-bc-1) \left(|d|^2-|b|^2\right)^{-2\sigma_1}\times\left(|c|^2-|a|^2\right)^{-2\sigma_2}\\
\end{equation}
This is essentially a simpler special case of the integral considered in appendix F of \cite{Harlow:2011ny}, hence most of the calculation is just a retread of the steps in there which for the sake self-containment are repeated here. First apply the change of integration variables of \cite{Zamolodchikov:1995aa} given by
\begin{equation}
\xi_1=\frac{b}{d}, \;\;\; \xi_2=\frac{a+b}{c+d}, \;\;\; \xi_3=\frac{a}{c}.
\end{equation}
This results in the integral
\begin{equation}
I(\sigma_1,\sigma_2)= \int \frac{d^2\xi_1 d^2\xi_2 d^2\xi_3}{|\xi_{12}|^{2-2(\sigma_1-\sigma_2)}|\xi_{23}|^{2-2(\sigma_2-\sigma_1)}|\xi_{31}|^{2-2(\sigma_1+\sigma_2)}}(1-|\xi_1|^2)^{-2\sigma_1}(1-|\xi_3|^2)^{-2\sigma_2},
\end{equation}
where $\xi_{ij}=\xi_i-\xi_j$. This integral can be simplified by exploiting the fact that $SU(1,1)$ transformations\footnote{unlike \cite{Zamolodchikov:1995aa,Harlow:2011ny} the group that leaves the integrand invariant is not $SU(2)$ in this case. Essentially because their light operators live on the sphere due to a lack of heavy operators. In contrast, our light operators are assumed to live on a space related by diffeomorphism to the Poincar\'e disk, i.e. our heavy operators are assumed to satisfy the Gauss-Bonnet bound discussed in section 3.}  of the form
\begin{equation}
\xi_i \rightarrow \frac{a\xi_i+b}{\bar{b}\xi_i +\bar{a}}, \;\; |a|^2-|b^2|=1,
\end{equation}
keep the integrand invariant. In particular, one can select an $SU(1,1)$ transformation that takes $\xi_1$ to infinity, resulting in the simplified expression
\begin{equation}
I(\sigma_1,\sigma_2)=\int d^2\xi_2 d^2\xi_3 \; |\xi_{23}|^{2(\sigma_2-\sigma_1)-2} (1-|\xi_3|^2)^{-2\sigma_2}.
\end{equation}
This integrand at least naively seems to converge for $\sigma_2>\sigma_1$. Of course, the final result should treat $\sigma_1$ and $\sigma_2$ on equal footing, the inequivalence is a relic resulting from the choice of sending $\xi_1$ to infinity. This integral can be rewritten into a quadruple integral over four real integration variables
\begin{equation}
I(\sigma_1,\sigma_2)=\pi\int_{-\infty}^{\infty} dx\, dy\, du\, dv\; \left((x-u)^2+(y-v)^2\right)^{\sigma_2-\sigma_1-1}\left(1-x^2-y^2\right)^{-2\sigma_2}
\end{equation}
After changing the integration variables $x,y,u,v \rightarrow ix,iy,iu,iv$, this integral can be recognized as a phase-shifted special case of an integral identity for the classical limit of the DOZZ formula \cite{Zamolodchikov:1995aa,Dorn:1994xn}\footnote{see equation F.9 of \cite{Harlow:2011ny}}. Hence we arrive at the conclusion
\begin{equation}
I(\sigma_1,\sigma_2)=2\pi^3 (-1)^{\sigma_2-\sigma_1-1}\lim_{\epsilon\rightarrow 0}\;  \epsilon \, \frac{\Gamma(\sigma_2-\sigma_1)\Gamma(\sigma_1-\sigma_2)\Gamma(\sigma_1+\sigma_2)\Gamma(\sigma_1+\sigma_2-1)}{\Gamma(2\sigma_1)\Gamma(2\sigma_2)}.
\end{equation}
We can see that for generic values of $\sigma_1$ and $\sigma_2$ the integral vanishes. The exception is when $\sigma_1=\sigma_2-\epsilon$. Hence in the body of the text it was stated that
\begin{equation}
I(\sigma_1,\sigma_2) \propto \delta(\sigma_1-\sigma_2).
\end{equation}
The integral also yields finite values whenever $\sigma_1$ and $\sigma_2$ are separated by integer values, we will not comment on this any further.

\FloatBarrier


\end{document}